\documentclass[journal,twoside,web]{ieeecolor}
\bstctlcite{IEEEexample:BSTcontrol}
\usepackage{generic}
\usepackage{cite}
\usepackage{amsmath,amssymb,amsfonts}
\usepackage{algorithmic}
\usepackage{graphicx}
\usepackage{textcomp}
\usepackage[caption=false]{subfig}
\def\BibTeX{{\rm B\kern-.05em{\sc i\kern-.025em b}\kern-.08em
		T\kern-.1667em\lower.7ex\hbox{E}\kern-.125emX}}
\begin{document}
	\title{Fractional Fowler-Nordheim Law for Field Emission from Rough Surface with Nonparabolic Energy Dispersion}
	\author{Muhammad Zubair, Yee Sin Ang, and Lay Kee Ang
		\thanks{Muhammad Zubair is currently affiliated with the Department of Electrical Engineering at Information Technology University of the Punjab, Ferozpur Road, 54000 Lahore, Pakistan. He was affiliated until recently with the SUTD-MIT International Design Center, Singapore University of Technology and Design, 487372 Singapore (e-mail: muhammad.zubair@itu.edu.pk, m.zubair.engr@gmail.com).}
		\thanks{Yee Sin Ang, and Lay Kee Ang are affiliated with the SUTD-MIT International Design Center, Singapore University of Technology and Design, 487372 Singapore (e-mails: yee\_sin@sutd.edu.sg, ricky\_ang@sutd.edu.sg). }	
		\thanks{Authors are sponsored by USA AFOSR AOARD (FA2386-14-1-4020), Singapore Temasek Laboratories (IGDS S16 02 05 1) and A*STAR IRG (A1783c0011).}
	}
	\maketitle

\begin{abstract}
The theories of field electron emission from perfectly planar and smooth canonical surfaces are well understood, but they are not suitable for describing emission from rough, irregular surfaces arising in modern nanoscale electron sources. Moreover, the existing models rely on Sommerfeld's free-electron theory for the description of electronic distribution which is not a valid assumption for modern materials with nonparabolic energy dispersion. In this paper, we derive analytically a generalized Fowler-Nordheim (FN) type equation that takes into account the reduced space-dimensionality seen by the quantum mechanically tunneling electron at a rough, irregular emission surface. We also consider the effects of non-parabolic energy dispersion on field-emission from narrow-gap semiconductors and few-layer graphene using Kane's band model. The traditional FN equation is shown to be a limiting case of our model in the limit of a perfectly flat surface of a material with parabolic dispersion. The fractional-dimension parameter used in this model can be experimentally calculated from appropriate current-voltage data plot. By applying this model to experimental data, the standard field-emission parameters can be deduced with better accuracy than by using the conventional FN equation.  


\end{abstract}

\maketitle


\section{\label{sec:Introduction} Introduction}

The field electron emission from the surface of a material is a well-known quantum mechanical process which has found a vast number of technological applications including field-emission display (FEDs), electron microscopes, electron guns and vacuum nanoelectronics~\cite{egorov2017field}. Fowler-Nordheim (FN) theory describes the electric field induced electron tunneling from a flat perfectly conducing planar surface through an approximately triangular potential-energy barrier~\cite{fowler1928electron,liang2013quantum}. This leads to a FN-type equation given here as

 \begin{eqnarray}
 J_{FN}= a_{FN} \frac{F^{2}}{\phi}\exp\left(-\frac{\nu b_{FN}\phi^{3/2}}{F}\right).
 \label{eqn:FN_origina}
 \end{eqnarray}
 
This gives the local emission current density $J_{FN}$ in terms of local work-function $\phi$ and surface electric field $F$. The symbols $a_{FN}(\approx 1.541434 \mu eVV^{-2}$)  and $b_{FN} (\approx 6.830890eV^{-3/2}Vnm^{-1})$ denote first and second FN constants~\cite{liang2013quantum}, and $\nu$ is a correction factor associated with the barrier shape. For exactly triangular barrier we take $\nu=1$. Taking the tunneling barrier as exactly triangular is not always physically realistic, so the barrier seen by tunneling electron is conventionally modeled as image-force-rounded model barrier known as ``Schottky-Nordheim (SN)" barrier~\cite{liang2013quantum}. Typically, the barrier shape correction factor ($\nu_{SN}$) has a value of 0.7 which enhances the predicted current density by approximately two orders of magnitude~\cite{forbes2006simple}. However, later mathematical developments~\cite{forbes2007reformulation} have yielded more accurate approximation for $\nu_{SN}$. 

There have been many analytical or semi-analytical works so far that address emission from specific geometry emitters. For instance, the field-emission from sharp-tip~\cite{sun2013analysis}, spherical~\cite{holgate2017field} and hyperbolic~\cite{jensen2017current} emitters have also been studied recently. At present, the situation can be viewed from a different standpoint. A mathematical treatment for slightly irregular or rough planar surface is required for better understanding of surface topographical effects on field-emission. Also, the original FN equation assumes quasi-free electron with parabolic energy dispersion. This assumption is no longer correct for materials with nonparabolic energy dispersion such as narrow-gap semiconductor and graphene monolayer~\cite{ang2017theoretical,ang2016current,ang2017relativistic}. 


Motivated by above, the purpose of this work is three-fold. 

In the first place, we re-derive the FN equation using a fractional-dimensional-space approach~\cite{stillinger1977axiomatic,zubair2012electromagnetic} by assuming that the effect of surface irregularity on the tunneling probability is captured by the space fractional-dimension parameter ($\alpha$). The underlying approach relies on fractional dimensional system of spatial coordinates to be used as an effective description of complex and confined systems (see~\cite{zubair2016fractional} and references therein for details). Some successful applications of this approach are in the areas of quantum field
theory~\cite{palmer2004equations},
general relativity~\cite{sadallah2009solution},
thermodynamics~\cite{tarasov2016heat},
mechanics~\cite{ostoja2014fractal},
hydrodynamics~\cite{balankin2012map},
electrodynamics~\cite{zubair2012electromagnetic,mughal2011fractional,naqvi2016cylindrical, zubair2011exact, asad2012electromagnetic, asad2012reflection,zubair2011exact2,zubair2011differential,zubair2011electromagnetic,zubair2010wave},
thermal transport~\cite{godinez2016space},
mechanics of anisotropic fractal materials~\cite{balankin2015continuum1}, and high current emission from rough
cathode~\cite{zubair2016fractional}.

Second, we compare the presented generalized FN theory with the conventional FN equations, showing its advantages. Based on the conventional FN equations, many experiments (some examples listed in~\cite{forbes2013development}) have reported field-enhancement factor (FEF) in the range of few thousands to over hundred thousands. Our model proposes a new field-enhancement correction factor which solves the issue of such spuriously high FEFs.
The historical FN theory~\cite{fowler1928electron} suggests that the empirical current-voltage ($i-V$) relationships of field emission should obey $i=CV^{k} \exp({-B/V})$, where $B$, $C$ and $\alpha$ are effectively constants and $k=2$. However, later developments~\cite{forbes2009use} showed that the values of $k$ may differ from 2 due to voltage dependence of various parameters in FN equations under non-ideal conditions. Based on our model, we propose that the experimental $i-V$ relationships of field emission from practical rough surfaces should approximately obey the empirical law $i=CV^{2\alpha} \exp({-B/V^{\alpha}})$, where $0 <\alpha\leq 1$. Since, $lim_{\frac{1}{V}\to \min} \frac{d \ln(i/V^2)}{d \ln(V)}=2\alpha-2$, the fractional parameter $\alpha$ for a given emission surface can be experimentally established from sufficiently accurate $i-V$ measurements. The knowledge of experimental $\alpha$ values would lead to a better physical understanding of field emission and its variations due to surface irregularities and material properties. It is emphasized that the feasibility of such measurements should be carefully explored by the experimentalists. 

Finally, using Kane's nonparabolic dispersion~\cite{askerov1994electron}, we also generalize the fractional FN theory for materials with highly nonparabolic energy dispersions. This extension will be important to model the field emission from rough surfaces of materials like narrow-gap semiconductors and Graphene, where surface morphology is expected to affect the emission properties. 



\section{\label{sec: Formulation} Fractional generalization of Fowler-Nordheim field emission equation}

Field emission involves the extraction of electrons from a solid by tunneling through the surface potential barrier. The Fowler-Nordheim method expresses the emission current density ($J_{FN}$) in terms of the product of supply function $N(E_\perp)$ and transmission coefficient $D(E_\perp)$ as following

\begin{eqnarray}
J_{FN}=\int_{0}^{\infty} N(E_\perp)D(E_\perp) dE_\perp.
\label{eqn:FN_main}
\end{eqnarray}
where $E_\perp$ is the normal energy, measured relative to the bottom of the conduction band. From free-electron theory, the supply function $N(E_\perp)$ is given as 
\begin{eqnarray}
N(E_\perp)= \frac{em_{e}}{2 \pi^2 \hbar^3} \ln \left[1+\exp\left(-\frac{E_\perp-E_{F}}{k_{B}T}\right)\right],
\label{eqn:supply_function}
\end{eqnarray}
where $k_{B}$ is Boltzmann's constant, $T$ is temperature, $\hbar$ is reduced Plank's constant, and $E_{F}$ is the Fermi energy. It is to be noted that this form of supply function is valid only for conducting emitters, and ignores the quantum confinement effects which may arise in nanoscale wire emitters~\cite{qin2011analytical}. For narrow gap semiconductors and Dirac materials, we need other forms of supply function due to the anisotropic distributions of electrons~\cite{liang2015electron,ang2017theoretical} (see section (\ref{sec: Formulation2}) for more details). It is known that, in general, the Eq. (\ref{eqn:supply_function}) provides an adequate description of the supply function for varying shape of emitters as the effect of emitter shape is included in external potential barrier only~\cite{holgate2017field}.  

In quantum mechanics, the transmission coefficient through a potential barrier is defined by $D(E_\perp)= j_{t}/j_{i}$, where transmitted, incident ($j_{t}$,$j_{i}$) probability current density is related to respective wave functions ($\psi_{t}$,$\psi_{i}$) by $j_{t,i}=\frac{i \hbar}{2m_{e}} \left(\psi_{t,i} \frac{\partial}{\partial x} \psi_{t,i}^{*}-\psi_{t,i} ^{*}\frac{\partial}{\partial x} \psi_{t,i}\right)$

FN's elementary theory uses the physical simplification
that tunneling takes place from a flat planar surface, through an exact triangular (ET) barrier given by $V_{b}(x)=E_{0}-eFx$, where $E_{0}=E_{F}+\phi$ and $\phi$ is the work function  at zero bias ($F=0$). Thus, the Schr\"{o}dinger equation in $x$ direction 
  
\begin{eqnarray}
&&-\frac{\hbar^2}{2 m_{e}}\frac{\partial^2 \psi_{i}(x)}{\partial x^2}=E_\perp\psi_{i}(x),\quad x\leq0,\\
&&-\frac{\hbar^2}{2 m_{e}}\frac{\partial^2 \psi_{t}(x)}{\partial x^2}+V_{b}(x)\psi_{t}(x)=E_\perp\psi_{t}(x),~x \geq 0.
\label{eqn:Schrodinger_integer}
\end{eqnarray}

\begin{figure}[!ht]
	\centering
	\includegraphics[width=0.45\textwidth]{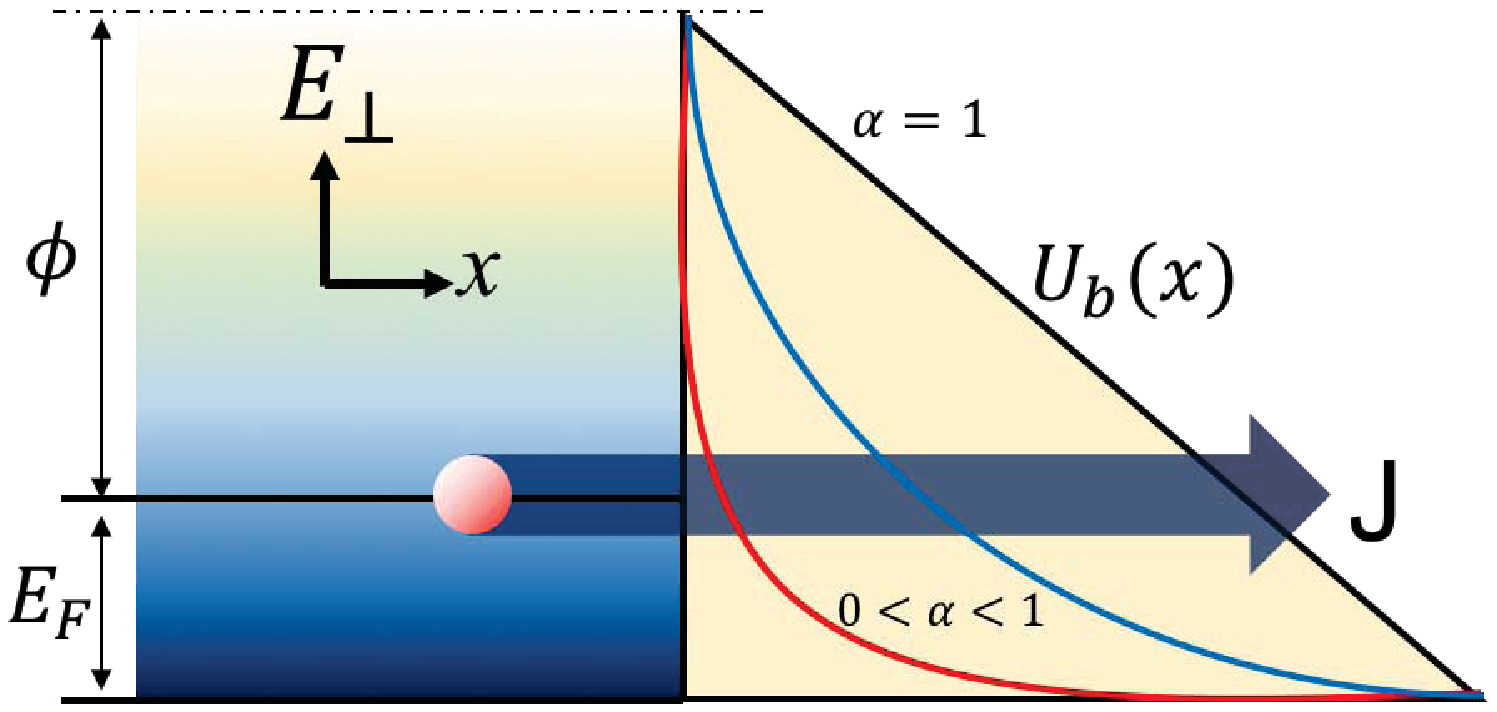}
	\caption{Schematic diagram for fractional generalization of field emission in solids. Here, $\alpha=1$ corresponds to the exactly triangular (ET) barrier at perfectly smooth planar interface, while $0<\alpha\leq1$ corresponds to reduced-dimensionality seen by the tunneling electrons at rough surface. The shape of barrier energy potential $U_{b}(x)$ varies with parameter $\alpha$. }
	\label{fig:FE}
\end{figure}

From Wentzel-Kramers-Brillouin (WKB) solution~\cite{liang2013quantum}, this standard formulation leads to the FN equation given in (\ref{eqn:FN_origina}). 
However, the effect of irregular, anisotropic solid surface can be effectively modeled by assuming that the potential barrier lies in a fractional-dimensional space with dimension $0 < \alpha \leq 1$ (see Fig.~\ref{fig:FE}). In doing so, the effect of roughness is accounted instead of traditional approach of assuming $F=\beta_{app} F_{0}$ in Eq.~(\ref{eqn:FN_origina}), where $\beta_{app}$ is the apparent FEF and $F_{0}$ is macroscopic field. Hence, the Eq. (\ref{eqn:Schrodinger_integer}) can be replaced with fractional-dimensional Schr\"{o}dinger equation given by 
  
\begin{eqnarray}
-\frac{\hbar^2}{2 m_{e}} \nabla_{\alpha}^2\psi_{t}(x)+V_{b}(x)\psi_{t}(x)=E_\perp \psi_{t}(x),\qquad x \geq 0,
\label{eqn:FD_schrodinger}
\end{eqnarray}
where, $\nabla_{\alpha}^2$ is the modified Laplacian operator in fractional-dimensional space given by~\cite{tarasov2014anisotropic,zubair2016fractional}
\begin{eqnarray}
\nabla_{\alpha}^2=\frac{1}{c^2(\alpha,x)}\left(
\frac{\partial^2}{\partial
	x^2}-\frac{\alpha-1}{x}\frac{\partial}{\partial
	x}\right),
\label{eqn:nabla_D}
\end{eqnarray}
with 
\begin{eqnarray}
c(\alpha,x)=
\frac{\pi^{\alpha/2}}{\Gamma(\alpha/2)}|x|^{\alpha-1}.
\label{eqn:dos1}
\end{eqnarray}

It can be seen that the effect of the non-integer (fractional) dimensions in Eq. (\ref{eqn:FD_schrodinger})  is to modify the wave function and hence the probability current density so that it takes into account the measure distribution of the space. This approach can effectively model the non-ideal emission surfaces by considering the space dimension $\alpha$ as description of complexity, assuming $\alpha=1$ represents an ideal planar surface. From solution of Eq. (\ref{eqn:FD_schrodinger}), we can calculate the new form of transmission coefficient $D(E_\perp,\alpha)$ which will represent the tunneling probability and hence the emission current as a function of space dimension $\alpha$. 
We can write Eq. (\ref{eqn:FD_schrodinger}) as 
\begin{eqnarray}
&&\left[\frac{\partial^2}{\partial
	x^2}-\frac{\alpha-1}{x}\frac{\partial}{\partial
	x}-\frac{2m_{e}}{\hbar^2}f^2(\alpha)x^{2\alpha-2}\left(V_b(x)-E_\perp\right)\right]\nonumber\\&&\psi_{t}(x)=0,
\label{eqn:rearranged_FD_Schrodinger}
\end{eqnarray}
where, $f(\alpha)=\frac{\pi^{\alpha/2}}{\Gamma(\alpha/2)}$. 

Eq. (\ref{eqn:rearranged_FD_Schrodinger}), through substitution of $\psi_{t}(x)=\xi_{t}(x) x^{(\frac{\alpha-1}{2})}$, can be reduced to standard WKB form~\cite{meissen2013integral} given by
\begin{eqnarray}
\left[\frac{\partial^2}{\partial
	x^2}-\frac{2m_{e}}{\hbar^2}\left( U_{b}(x)-f^2(\alpha)x^{2\alpha-2}E_\perp \right)\right]\xi_{t}(x)\nonumber\\=0,
\label{eqn:comparable}
\end{eqnarray}
where, 
\begin{eqnarray}
 U_{b}(x)&&= -\frac{\hbar^2}{2 m_{e}} \left(\frac{\alpha-1}{2x^2}+\frac{(\alpha-1)^2}{4x^2}\right)+f^2(\alpha)x^{2\alpha-2}V_b(x) \nonumber\\ &&\approx f^2(\alpha)x^{2\alpha-2}V_b(x).
\label{eqn:Ub}
\end{eqnarray}

\begin{figure}[!ht]
	\centering
	\includegraphics[width=0.45\textwidth]{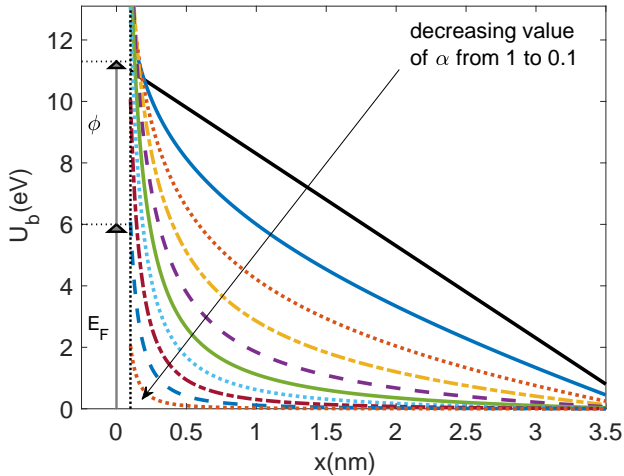}
	\caption{The exactly triangular and fractional potential energy barriers with varying fractional-dimension $\alpha$ for $\phi=5.3eV$, $E_{F}=6eV$, and $F=3eV/m$. $\alpha=1$ corresponds to the exactly triangular (ET) barrier.}
	\label{fig:potentials}
\end{figure}

The shape of this effective fractional potential-barrier in above equation is shown in Fig. (\ref{fig:potentials}) together with the exactly triangular barrier (corresponding $\alpha=1$).

By solving Eq. (\ref{eqn:comparable}) through WKB method, and matching the wave function~\cite{liang2013quantum}, we get the transmission coefficient of this barrier as

\begin{eqnarray}
D(E_\perp)=\exp \left(-g_{e} \int_{x1}^{x2}\sqrt{V_{0}(x)} dx\right) ,
\label{eqn:Tex}
\end{eqnarray}
with $V_{0}(x)=f^2(\alpha)x^{2\alpha-2}(V_{h}-eFx)$, where $V_{h}=\phi+E_{F}-E_\perp$ is the zero-field height of the potential barrier, $g_{e}=2\sqrt(\frac{2m_{e}}{\hbar^2})$, $x_{1}=0$ and $x_{2}=V_{h}/eF$. Using the binomial expansion and taking $eFx/V_{h}\ll1$, the above integral is approximated as
 \begin{eqnarray}
\int_{x1}^{x2}\sqrt{V_{0}(x)} dx&&=f(\alpha)\int_{x1}^{x2}x^{\alpha-1}\sqrt{(V_{h}-eFx)} dx\nonumber\\\approx&& 0.44 f(\alpha)\left(\frac{\alpha+2}{\alpha^2+\alpha}\right)\frac{V_{h}^{\alpha+1/2}}{(eF)^\alpha}.
\label{eqn:Integral}
\end{eqnarray}

Inserting Eq. (\ref{eqn:Integral}) into Eq. (\ref{eqn:Tex}), and following proper algebraic steps, we get
\begin{eqnarray}
D(E_\perp)\approx D_{F\alpha}\exp\left[\frac{E_\perp-E_{F}}{d_{F\alpha}}\right],
\label{eqn:Tex_final}
\end{eqnarray}
where,
\begin{eqnarray}
\label{eqn:D_alpha}
D_{F\alpha}=\exp\left[-b_{\alpha}\frac{\phi^{\alpha+1/2}}{F^\alpha}\right],\\
b_{\alpha}=0.44 f(\alpha) \left(\frac{\alpha+2}{e^\alpha(\alpha^2+\alpha)}\right)g_{e},
\label{eqn:balpha}
\end{eqnarray}
\begin{eqnarray}
\frac{1}{d_{F\alpha}}=0.22 f(\alpha) \left(\frac{2\alpha^2+5\alpha+2}{\alpha^2+\alpha}\right)\frac {g_{e}\phi^{\alpha-1/2}}{(eF)^\alpha},
\label{eqn:dalpha}
\end{eqnarray}
where $\phi$ is measured in $eV$ and $F$ in $V/nm$. The transmission coefficient depends on work function ($\phi$), electric field ($F$) and the fractional dimension parameter ($\alpha$). 

Substituting Eq. (\ref{eqn:supply_function}) and Eq. (\ref{eqn:Tex_final}) into Eq. (\ref{eqn:FN_main}), the fractional form of emission current density can be written as

\begin{eqnarray}
J_{FN\alpha}= &&\frac{em_{e}}{2 \pi^2 \hbar^3}  D_{F\alpha} 
\int_{0}^{\infty} \exp\left[\frac{E_\perp-E_{F}}{d_{F\alpha}}\right] \nonumber\\ &&\times\ln \left[1+\exp\left(-\frac{E_\perp-E_{F}}{k_{B}T}\right)\right]dE_\perp.
\label{eqn:FN_frac1}
\end{eqnarray}

At room temperature, where electrons near Fermi energy dominate the tunneling, we can use approximation $k_{B}T \ln \left[1+\exp\left(-\frac{E_\perp-E_{F}}{k_{B}T}\right)\right]=E_{F}-E_\perp$, to get

\begin{eqnarray}
J_{FN\alpha}= &&\frac{em_{e}}{2 \pi^2 \hbar^3}  D_{F\alpha}\nonumber\\ &&\times
\int_{0}^{E_{F}} \exp\left[\frac{E_\perp-E_{F}}{d_{F\alpha}}\right] (E_{F}-E_\perp)dE_\perp.
\label{eqn:FN_frac2}
\end{eqnarray}

In solving Eq.~(\ref{eqn:FN_frac2}), the generalized fractional field-emission density can be given in the FN form as 
 
 \begin{eqnarray}
 J_{FN\alpha}= a_{FN\alpha} \frac{F^{2\alpha}}{\phi^{2\alpha-1}}\exp\left(-\frac{b_{FN\alpha}\phi^{\alpha+1/2}}{F^\alpha}\right),
 \label{eqn:FN_frac_final}
 \end{eqnarray}
where 

 \begin{eqnarray}
 a_{FN\alpha}= \frac{e^{2\alpha+1}m_{e}}{2 \pi^2 \hbar^3 g_{e}^2} \frac{1}{0.0484 f^2(\alpha)} \nonumber\\ \times \left(\frac{(\alpha^2+\alpha)^2}{(2\alpha^2+5\alpha+2)^2}\right),
 \label{eqn:aFN_alpha}
 \end{eqnarray}
 \begin{eqnarray}
 b_{FN\alpha}=0.44 f(\alpha) \left(\frac{\alpha+2}{e^\alpha(\alpha^2+\alpha)}\right)g_{e}.
 \label{eqn:balpha_FN}
 \end{eqnarray}
 
 Notice that for $\alpha=1$ (ideal planar emission surface), the Eq. (\ref{eqn:FN_frac_final}-\ref{eqn:balpha_FN}) reduce exactly to the original FN equation (\ref{eqn:FN_origina}) and the standard FN constants, respectively.  It is important to establish a correct value of parameter $\alpha$ while fitting the experimental data in order to extract the standard field-emission parameters like FEF and emission area. 
 
 In the traditional approach, $F=\beta_{app} F_{0}$ is assumed in the exponential part of Eq.~(\ref{eqn:FN_origina}), where $\beta_{app}$ is the apparent FEF, $F_{0}$ is macroscopic field and $F$ is the surface field of the emitter. This $\beta_{app}$ is chosen by using the slope ($S_{fit}$) of experimental FN ($J_{FN}/F^2$ versus $1/F$) plot in the expression $\beta_{app}=-b_{FN}\phi^{3/2}/S_{fit}$~\cite{forbes2009use}. However, in many experiments, spuriously high values of $\beta_{app}$ have been reported (see~\cite{forbes2013development} for examples) ranging from few thousands to above hundred thousand in some cases, and are less comprehensible if regarded as values for true electrostatic FEFs. Our model of Eq. (\ref{eqn:FN_frac_final}) suggests that such spurious values of FEF could result from the traditional assumption of $\alpha=1$, which is not true for irregular, rough surfaces. Taking $F=\beta_{app} F_{0}$ in Eq. (\ref{eqn:FN_frac_final}) and comparing with original FN Eq.~(\ref{eqn:FN_origina}), we can see that the actual value of FEF (called $\beta_{actual}$) may differ from $\beta_{app}$ for varying values of $\alpha$ and can be given written as 
 
  \begin{eqnarray}
\beta_{actual}=\beta_{corr}\beta_{app}
  \label{eqn:beta_actual}
  \end{eqnarray}
 
 where, the correction factor can be written as
 
  \begin{eqnarray}
  \beta_{corr}=\frac{b_{FN}F_{0}^{\alpha-1}}{b_{FN\alpha}\phi^{\alpha-1}}
  \label{eqn:beta_corr}
  \end{eqnarray}
 
\begin{figure}[!ht]
 	\centering
 	\includegraphics[width=0.45\textwidth]{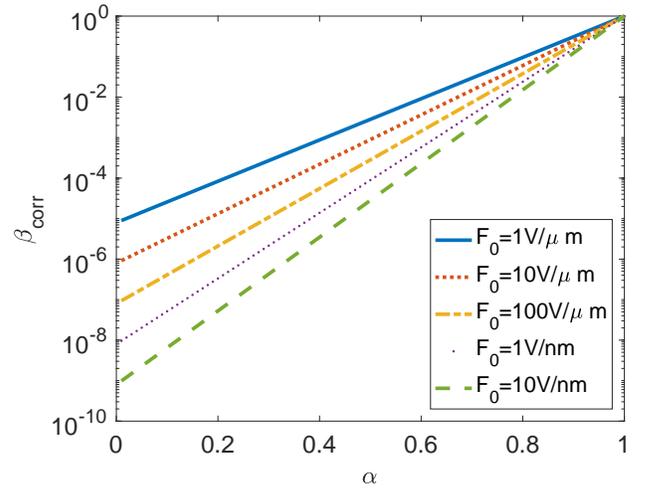}
 	\caption{Field-enhancement correction factor ($\beta_{corr}$) versus fractional-dimension parameter $\alpha$ at varying values of $F$ and $\phi=5.3eV$ using equation (\ref{eqn:beta_corr})}
 	\label{fig:beta_corr_factor}
 \end{figure}

 Fig. (\ref{fig:beta_corr_factor}) shows field-enhancement correction factor ($\beta_{corr}$) versus fractional-dimension parameter $\alpha$ at varying values of $F$ and $\phi=5.3eV$. At $\alpha=1$, the correction factor is unity for perfectly smooth planar surface. It is shown in analytical calculations for various cathode geometries that the electrostatic FEF varies from unity to a few hundred~\cite{zhu2015space,lin2017electric,zhu2013novel,sun2012onset}. However, in many reported experiments (see~\cite{forbes2013development} for examples), the reported FEFs ($\beta_{app}$) have spuriously high values. In practical applications, it is assumed that $\beta_{actual}$ will always be greater than 1 up to a few hundred as suggested by electrostatic field calculations at various geometries. This means that there is always a lower bound on $\beta_{corr}$ depending on $\beta_{app}$ and correctly calculated values of $\alpha$. Using this correction factor ($\beta_{corr}$), the spuriously high values of FEF can be corrected, given that the parameter $\alpha$ is correctly established using the procedure discussed at the end of this section. It is to be noted that the $\beta_{corr}$ takes into account the voltage or field dependence of FEF through $F_{0}^{\alpha-1}$ term. Such voltage dependent FEF has also been studied recently~\cite{forbes2017theoretical}.

 Now, assuming $F=\gamma V$, where $\gamma $ is a constant, and that the emission area $A$ is constant in the relation $i=AJ_{FN}$, the Eq. (\ref{eqn:FN_origina}) suggests that the empirical field-emission $i-V$ characteristics should obey 
 \begin{equation}
 i=CV^{k}\exp(-B/V)
 \label{eqn:FN_Forbes_emperical}
 \end{equation}
 where, $B$, $C$ and $k$ are constants and $k=2$. Due to Eq. (\ref{eqn:FN_Forbes_emperical}), it is modern customary experimental practice to plot field-emission results as an FN-plot (i.e., $\ln (i/V^k)$ versus $1/V$), assuming that this should generate an exact straight line for $k=2$. The 1928 FN theory assumes an exact triangular tunneling barrier, which is not physically realistic for more practical problems where emission area $A$ and barrier-shape correction factor $\nu$ may become field dependent~\cite{forbes2008call}. Thus, it is believed that the true value of $k$ in Eq. (\ref{eqn:FN_Forbes_emperical}) might be different than 2. Several experimentalist have tried fitting their experimental data with arbitrary values of $k$ ranging from -1 to 4, usually with a goal to achieve straight-line FN plot. A complete history of various attempts to find the correct value of $k$ has been summarized in~\cite{forbes2009use}. Some of these results had no obvious theoretical explanation and have largely been ignored including $k=4$ suggested by Abbot et. al., ~\cite{abbott1939range} and $k=0.25$ by Oppenheimer~\cite{oppenheimer1928quantum}. 
 
 The most notable theoretical explanation for $k\neq2$ is based on approximate calculation of SN barrier-shape correction factor $\nu_{SN}$ by Forbes (see~\cite{forbes2008call} for details) which suggests $k=2-\eta/6$, where $\eta=b_{FN}\phi^{3/2}/F_{\phi}$ with $F_{\phi}= (4\pi \epsilon_{0}/e^3) \phi^2$. Hence, the value of $k$ varies for different values of work-function $\phi$. For $\phi=4.5 eV$, we get $F_{\phi}=14$ V/nm,  $\eta=4.64$, and  $k=2-\eta/6=1.2$~\cite{forbes2008call}. The main factor that determines the value of $k$ so far is the explicit field dependence that appears in the pre-exponential part of FN equation. However, the shape of the shape of potential-barrier and field enhancement at surface of the emitter may also effect the exponential part in FN equation.

 In our reported model, we have modeled these realistic effects in terms of ``effective space dimensionality" through which the electrons move prior to emission. From our results in Eq. (\ref{eqn:FN_frac_final}), we suggest a more generalized form of empirical law given as
 \begin{equation}
 i=CV^{2\alpha}\exp(-B/V^\alpha)
 \label{eqn:FN_Frac_emperical}
 \end{equation}
 where, $\alpha$ is fractional space dimensionality parameter with $0<\alpha\leq1$. Eq. (\ref{eqn:FN_Frac_emperical}) is clearly more generalized form of Eq. (\ref{eqn:FN_Forbes_emperical}) with an addition parameter in the exponential term which can capture the field-enhancement effect of non-planar surfaces. This suggests a fitting of experimental data with Eq. (\ref{eqn:FN_Frac_emperical}), taking $\alpha$ as an unknown.  One could try plotting $\ln (i/V^{2\alpha})$  versus $1/V^{\alpha}$ in search of a straight-line plot, but, this is not a good method when $\alpha$ is unknown. Alternatively, from Eq. (\ref{eqn:FN_Frac_emperical}) 
 
 \begin{equation}
 \ln(i/V^2)=\ln(C)+(2\alpha-2)\ln(V)-B/V^\alpha,
 \label{eqn:FN_Frac_emperical_simpl1}
 \end{equation}
 \begin{equation}
 \frac{d \ln(i/V^2)}{d \ln(V)}=2\alpha-2+\alpha B/V^\alpha.
 \label{eqn:FN_Frac_emperical_simpl2}
 \end{equation}
 
 \begin{equation}
 lim_{{\frac{1}{V}\to \min}} \frac{d \ln(i/V^2)}{d \ln(V)}=2\alpha-2.
 \label{eqn:FN_Frac_emperical_simpl3}
 \end{equation}
 
 \begin{figure}[!ht]
 	\centering
 	\subfloat[\label{fig:enhancementxo}]{%
 		\includegraphics[width=.24\textwidth]{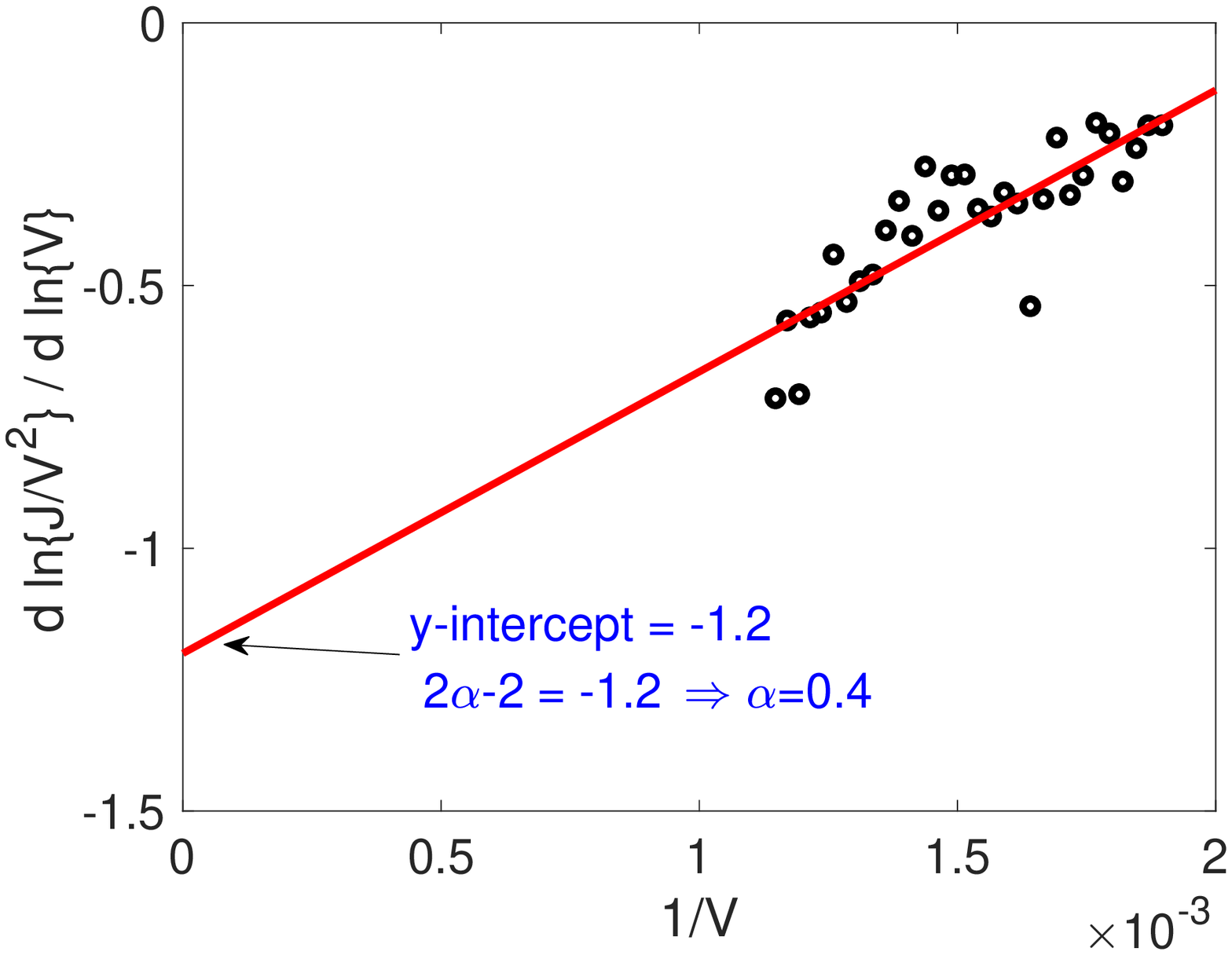}%
 		\label{Fig:Alpha0x4}
 	}
 	\subfloat[\label{fig:enhancementalpha}]{%
 		\includegraphics[width=.24\textwidth]{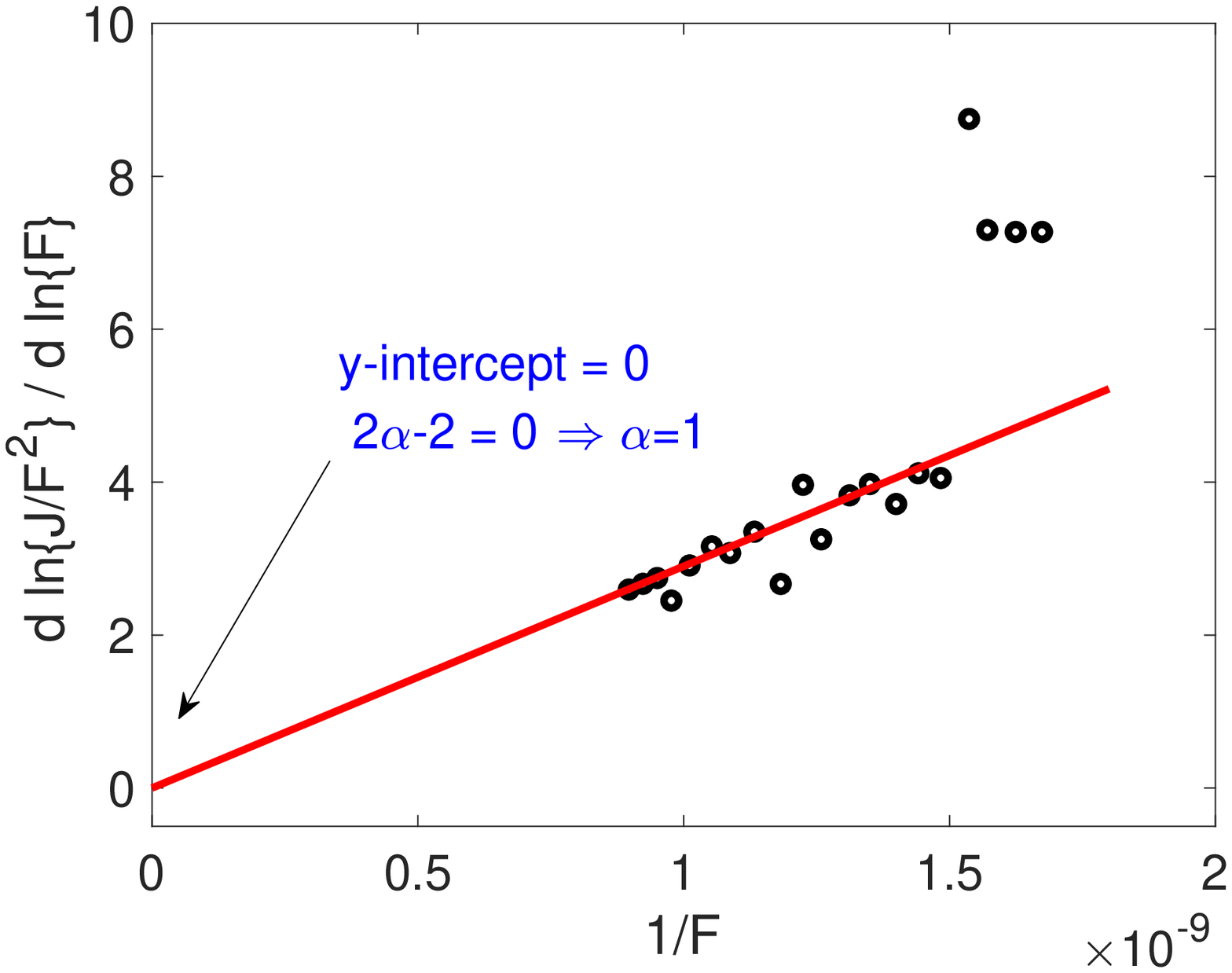}%
 		\label{Fig:Alpha1}
 	} 	
 	\hfill
 	\subfloat[\label{fig:enhancementxo}]{%
 		\includegraphics[width=0.24\textwidth]{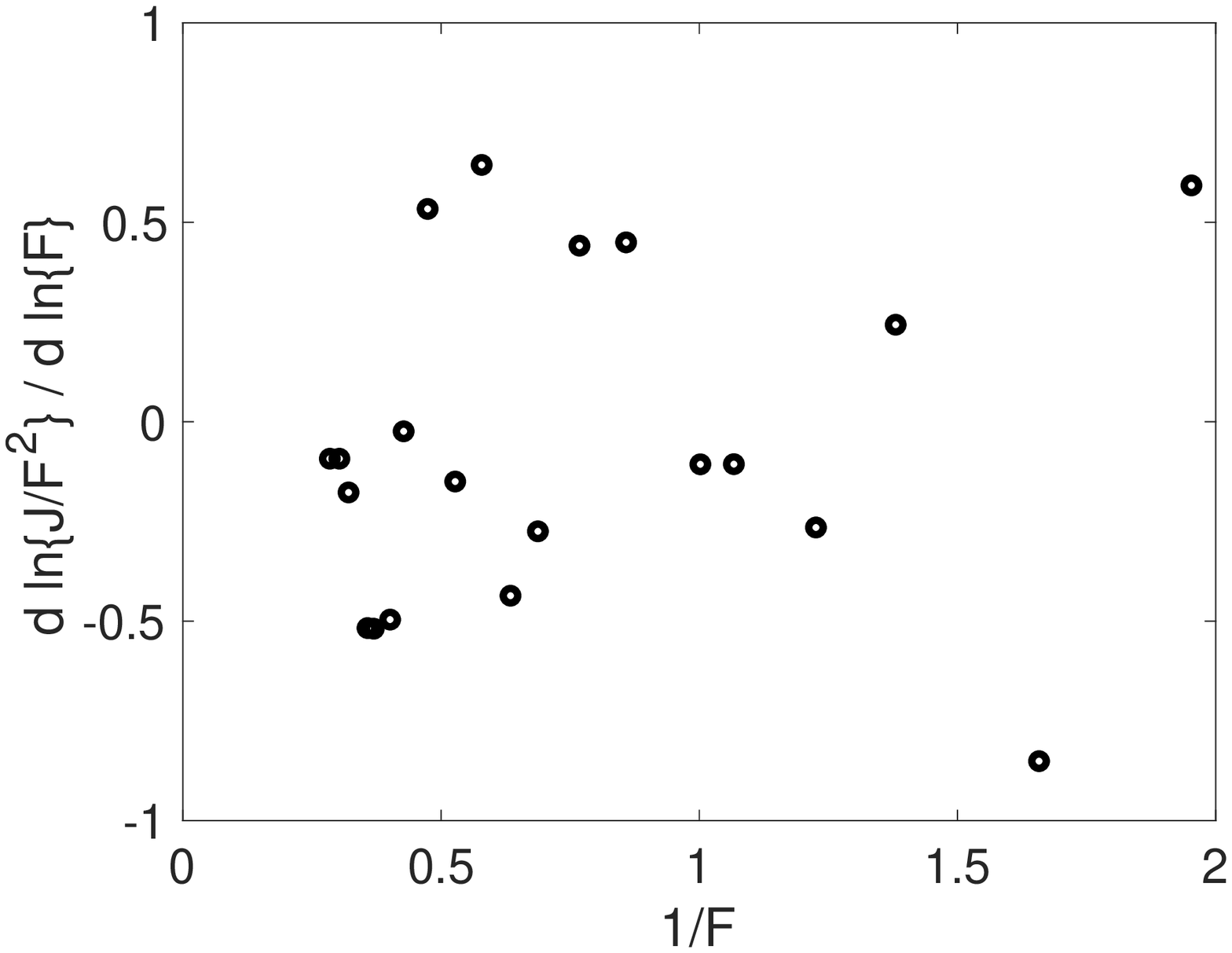}%
 		\label{Fig:AlphaNoisy}
 	}
 	\caption{Extraction of fractional-dimension parameter $\alpha$ from appropriate current-voltage or current-field plots (a) data set for field-emission of ZNO nanopencils taken from ~\cite{wang2005zno} (b)
 		data set for field-emission of carbon nanotubes from~\cite{li2017carbon} (c)
 		data set for field-emission of nanoshuttles taken from~\cite{li2011single}, here the required form of data is too noisy to extract the $\alpha$ parameter. We emphasize the need for direct measurement of $di/dV$ in order to correctly establish the $\alpha$ parameter.} \label{Fig:Extraction_of_alpha}
 \end{figure}

 Using Eq. (\ref{eqn:FN_Frac_emperical_simpl3}), the value of $\alpha$ can be calculated from y-intercept of ``$\frac{d \ln(i/V^2)}{d \ln(V)}$ versus $1/V$" or ``$\frac{d \ln(J/F^2)}{d \ln(F)}$ versus $1/F$" plot. This plotting method was tested on three experimental data sets shown in Fig. (\ref{Fig:Extraction_of_alpha}). In Fig. (\ref{Fig:Alpha0x4}), the field-emission data from Ref.~\cite{wang2005zno} was extracted and processed using the above mentioned procedure. It is shown that the extracted value of $\alpha$ is 0.4. This implies that classical FN equation (assuming $\alpha=1$) is not suitable to extract emission parameters in this case. Following the same procedure on data taken from Ref.~\cite{li2017carbon} we show that the extracted value of $\alpha$ parameter is 1, so the classical FN equation seems valid in this case. Finally, Fig. (\ref{Fig:AlphaNoisy}) shows the data processed from Ref.~\cite{li2011single}. In this case the resulting plot is too noisy for useful conclusions. It is clear that this method requires current-voltage data form carefully crafted experiments with minimal noise. The similar problem of noisy data has already been encountered in~\cite{forbes2008call}. An alternate form of Eq. (\ref{eqn:FN_Frac_emperical_simpl3}) would be better in which  ``$\frac{d \ln(i/V^2)}{d \ln(V)}=\frac{V}{i}\frac{di}{dV}-2$". The direct measurement of $di/dV$ would be more accurate by using some phase-sensitive detection techniques suggested in ~\cite{forbes2008call}. 
 
 It should be noted that the fractional FN model developed above does not take into account the space-charge-limited (SCL) effect, which is expected to become dominant at high bias regime. In a previous work~\cite{zubair2016fractional}, we show that the \emph{fractional Child-Langmuir} (CL) model predicts a stronger current for a rough surface when compared to the classical CL law. It is possible to combine the fractional FN law with the fractional CL law in a future work to study the continuous transition between the two transport regimes. Nonetheless, a conservative estimation can be obtained by calculating the \emph{transitional bias voltage}, $V_{T}$, at which the fractional FN current is equal to the fractional CL current. We found that $V_T$ is related with vacuum gap $D$ and parameter $\alpha$ through the relation $D^{\alpha}=\ln\left(\frac{9}{4\sqrt{2}}V_{T}^{(2\alpha-3/2)}\right) [V_{T}]^{\alpha}$. This $V_T$ serves as a conservative upper limit for the fractional FN law to be valid. For bias voltage larger than $V_T$, the SCL effect needs to be explicitly taken into account. We further note that, for the emerging class of novel two-dimensional (2D) materials, it is well-known that their surfaces often exhibit corrugations and crumples \cite{deng2016wrinkled}. The fractional approach developed here may also be extended to model the current-voltage characteristics of 2D-material-based planar field emitter, which has recently been shown to exhibit non-Fowler-Nordheim behavior and a complete absence of SCL effect \cite{ang2017generalized}.

 \section{ \label{sec: Formulation2} Derivation of field emission with Kane's nonparabolic dispersion}
 
 In this section, Kane's non-parabolic dispersion is used to derive a general field-emission model that covers both non-relativistic and relativistic charge carrier regimes. Such model was developed to describe the finite coupling between conduction and valence band in narrow-bandgap semiconductor and also in the high energy regime of semiconductor where band non-parabolicity becomes non-negligible. Kane's dispersion~\cite{askerov1994electron} can be written as
 \begin{equation}
 E_\parallel (1 + \gamma E_\parallel )=\frac{\hbar^2 k_\parallel^2}{2m^*} ,
 \label{eqn:YS1}
 \end{equation}
 where $\gamma$ is a non-parabolic parameter and $m^*$ is the electron effective mass. For $\gamma \to 0$, the conventional parabolic energy dispersion, $E_\parallel(\gamma\to0) = \hbar^2k_\parallel^2 / 2m^*$, is obtained. For $\gamma \to \infty$, Kane's dispersion leads to a $k_\parallel$-linear form of
 \begin{equation}
 E_\parallel (\gamma\to\infty) = \hbar v_F k_\parallel,
 \end{equation}
 where $v_F \equiv 1/\sqrt{2m^*\gamma}$. Eq. (\ref{eqn:YS1}) can be explicitly solved as
 \begin{equation}
 E_\parallel = \frac{\sqrt{1 + \frac{4\gamma \hbar^2 k_\parallel^2}{2m^*}} -1 }{2\gamma}.
  \label{eqn:YS3}
 \end{equation}
 By differentiating the left hand side of Eq. (\ref{eqn:YS1}) with respect to $k_\parallel$, we obtain the following relation
 \begin{equation}
 k_\parallel dk_\parallel = \frac{2m^*}{\hbar^2} \left(1 + 2\gamma E_\parallel\right) dE_\parallel.
   \label{eqn:YS4}
 \end{equation}
 Eq. (\ref{eqn:YS4}) shall play the critical role of determining the electron supply function density for field emission process.
 
 
 In general, the field emission current from a bulk material can be obtained by summing all $\mathbf{k}$-modes
 \begin{equation}
 J_{FN,Kane} = e g_{s,v}\sum_{\mathbf{k}} v(E_\perp)D(E_\perp) f(\mathbf{k}),
   \label{eqn:YS5}
 \end{equation}
 where $g_{s,v}$ is the spin-valley degeneracy, $v(E_\perp)$ and $D(E_\perp)$ are the carrier velocity and transmission probability along the emission direction, and $f(\mathbf{k})$ is the Fermi-Dirac distribution function. Here, it is assumed that the total energy of the carrier, $E(\mathbf{k})$, can be partitioned as $E(\mathbf{k}) = E_\parallel(\mathbf{k}_\parallel) + E_\perp(\mathbf{k}_\perp)$ where $\mathbf{k}_\parallel$ and $\mathbf{k}_\perp$ are the wave vectors transverse to and along the emission direction. In the continuum limit, Eq. (\ref{eqn:YS5}) becomes
 \begin{equation}
 J_{FN,Kane} = \frac{g_{s,v}e}{(2\pi)^3} \int d \mathbf{k}_\parallel \int v(E_\perp) D(E_\perp) f(\mathbf{k}) dk_\perp,
    \label{eqn:YS6}
 \end{equation}
 which can be simplified as followed: (i) the integral $\int_0^\infty (\cdots) v(E_\perp) dk_\perp $ can be replaced by $\hbar^{-1} \int_{0}^{\infty} (\cdots) dE_\perp$ since $v(E_\perp) = \hbar^{-1} dE_\perp /dk_\perp$; (ii) for field emission, only carriers up to the Fermi level takes part in the transport process, and thus the Fermi Dirac distribution can be represented by a two-variable Heaviside function, i.e. $f(\mathbf{k}) \to \Theta(E_\parallel + E_\perp - E_F) $ -- this further limits the $E_\perp$-and $E_\parallel$-integral to $\int_0^{E_F - E_\parallel} (\cdots)dE_\perp $ and $\int_0^{E_F} (\cdots)dE_\parallel$, respectively. Correspondingly, Eq. (\ref{eqn:YS6}) is simplified as
 \begin{equation}
 J_{FN,Kane}  = \frac{g_{s,v}e}{\hbar (2\pi)^2} \int_0^{E_F} k_\parallel dk_\parallel  \int_0^{E_F - E_\parallel} dE_\perp D(E_\perp).
   \label{eqn:YS7}
 \end{equation}
 By using the semiclassical WKB approximation of $D(E_\perp) = D_F \exp{ \left[ (E_\perp - E_F)/d_F \right]}$ for a triangular barrier, where $D_F$ and $d_F$ are constants defined in Eq. (\ref{eqn:D_alpha}) and (\ref{eqn:dalpha}) in the limit of $\alpha=1$, respectively, and taking into account Eq. (\ref{eqn:YS4}), Eq. (\ref{eqn:YS7}) becomes
 \begin{eqnarray}
 J_{FN,Kane}  &=& \frac{g_{s,v}e}{ (2\pi)^2} \frac{2m^*}{\hbar^3} D_F \int_0^{E_F} dE_\parallel (1 + 2\gamma E_\parallel) \nonumber\\ \times &&\int_0^{E_F - E_\parallel} e^{\frac{E_\perp - E_F}{d_F}} dE_\perp \nonumber \\
 &=& \frac{g_{s,v}e}{ (2\pi)^2} \frac{2m^*}{\hbar^3} D_F d_F \left[ A + B\right],
 \label{eqn:JFN_Kane}
 \end{eqnarray}
 where
 \begin{eqnarray}
 A &\equiv& \int_0^{E_F}dE_\parallel \left( e^{-\frac{E_\parallel}{d_F}} - e^{-\frac{E_F}{d_F}} \right) \nonumber \\
 &=& d_F \left[ 1 - \left( 1 + \frac{E_F}{d_F} \right) e^{-\frac{E_F}{d_F}} \right],
 \end{eqnarray}
 and 
 \begin{eqnarray}
 B &\equiv& 2\gamma \int_0^{E_F} dE_\parallel E_\parallel  \left( e^{-\frac{E_\parallel}{d_F}} - e^{-\frac{E_F}{d_F}} \right) \nonumber \\
 &=& 2\gamma d_F^2 \left\{ 1 - \left[ 1 + \frac{E_F}{d_F} + \frac{1}{2}\left(\frac{E_F}{d_F}\right)^2 \right]e^{-\frac{E_F}{d_F}} \right\}.
 \end{eqnarray}
 For typical conducting solid, $E_F \gg d_F$, hence, 
 \begin{equation}
 A \approx d_F,
 \end{equation}
 and
 \begin{equation}
 B \approx 2\gamma d_F^2, 
 \end{equation}
 From Eq. (\ref{eqn:JFN_Kane}), we obtained the field emission current density as
 \begin{equation}
 J_{FN,Kane}  \approx \frac{g_{s,v}e}{ (2\pi)^2} \frac{2m^*}{\hbar^3} D_F \left[ d_F^2 + 2\gamma d_F^3 \right].
 \label{eqn:JFN_Kane2}
 \end{equation}
 which is composed of two parts: a conventional FN component and a correction factor that accounts for band non-parabolicity.

 Using Eq. (\ref{eqn:Tex_final})  in Eq. (\ref{eqn:JFN_Kane}), the fractional generalization of Eq. (\ref{eqn:JFN_Kane2}) can be written as 
 \begin{equation}
 J_{FN\alpha,Kane}  \approx \frac{g_{s,v}e}{ (2\pi)^2} \frac{2m^*}{\hbar^3} D_{F\alpha} \left[ d_{F\alpha}^2 + 2\gamma d_{F\alpha}^3 \right].
 \label{eqn:FN_Kane_frac_final}
 \end{equation}
 Eq. (\ref{eqn:FN_Kane_frac_final}) is the most generalized from of Fowler-Nordheim field emission where band non-parabolicity is captured by the \textit{nonparabolicity parameter} ($\gamma$) and the interface irregularity, topography effect is captured by \textit{fractional dimension parameter} ($\alpha$).  Notice that for $\alpha=1$ (ideal planar emission surface) and $\gamma=0$ (conventional parabolic dispersion), the Eq. (\ref{eqn:FN_Kane_frac_final}) reduces exactly to the standard Fowler-Nordheim (FN) equation (\ref{eqn:FN_origina}). This fractinal generalization of FN theory for materials with highly non-parabolic dispersion will be very useful to study the field emission at surfaces of novel materials like narrow-gap semiconductors and Graphene.  
 


\section{\label{sec: Conclusions} Conclusions}

The standard FN theory based existing formulas for the field-induced emission of electrons from ideal planar surfaces are not always suitable for practical rough, irregular surfaces and may lead to significant order-of-magnitude errors when applied to experimental data incorrectly. Also, the standard models rely on free-electron theory assuming an isotropic distribution of electrons with parabolic energy dispersion. Such an assumption is no longer valid for narrow-gap semiconductors and few layer graphene where energy dispersion becomes non-parabolic due to band structure.

In this work, accurate equations for the field emission current densities from rough planar surfaces are derived using Kane's non-parabolic dispersion model. The reported expressions are applicable to metallic, narrow-gap semiconductor and few-layer Graphene planar electron emitters with rough surfaces in many modern applications. Based on our model, we have proposed a new form of field-enhancement correction factor $\beta_{corr}$ to avoid spuriously high FEFs in recent experiments~\cite{forbes2013development}. We propose that the experimental current-voltage ($i-V$) relationships of field emission from practical rough surfaces should approximately obey the empirical law $i=CV^{2\alpha} \exp({-B/V^{\alpha}})$, where $0 <\alpha\leq 1$. We have shown that the fractional-dimension parameter $\alpha $ can be established from the y-intercept of ``$\frac{d \ln(i/V^2)}{d \ln(V)}$ or $\frac{V}{i}\frac{di}{dV}-2$" versus ``$1/V$" plot. This requires an accurate phase-detection measurement of $\frac{di}{dV}$, previously proposed by Forbes in his work~\cite{forbes2008call}. Determination of $\alpha$ values could be very useful for better understanding of field emission and its variation as between different materials and surface roughness profiles. We strongly re-emphasize the Forbes's call of experiment to directly measure the $\frac{di}{dV}$ in field-emission experiments. The electronic availability of raw $i-V$ measurement data from experimentalists would be extremely useful for further explorations and improvements in the existing theoretical models. 

\bibliographystyle{IEEEtran}
\bibliography{Draft_FE}

\begin{IEEEbiography}[{\includegraphics[width=1in,height=1.25in,clip,keepaspectratio]{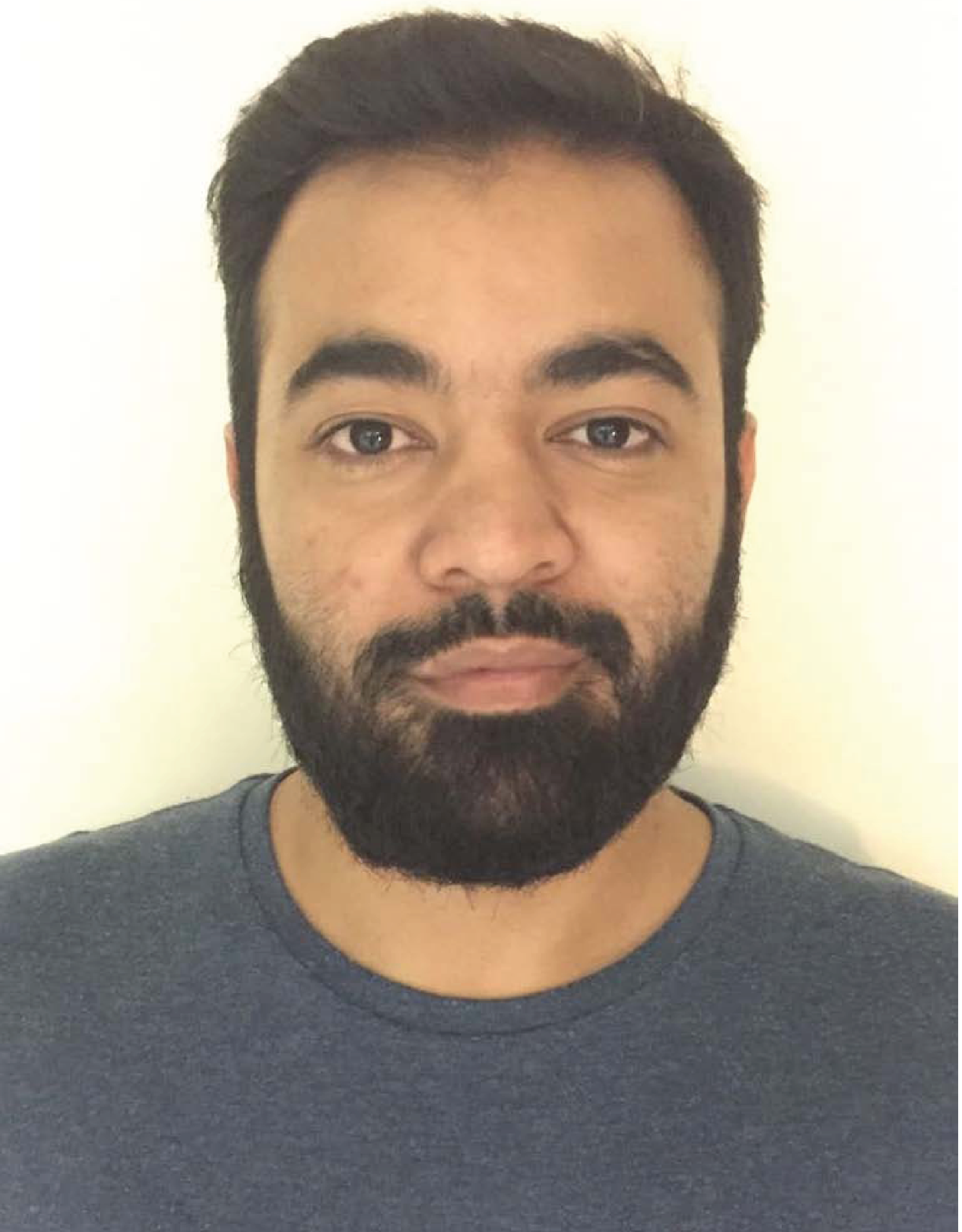}}]{Muhammad Zubair} received his PhD (Mar 2015) in electronic engineering at Politecnico di Torino, Italy. During his doctorate, he has received multiple advanced trainings on applied electromagnetism offered at European School of Antennas (ESOA). Part of his PhD research was completed at Istituto Superiore Mario Boella (ISMB) in collaboration with Boston University (BU). After his doctorate, he worked as a Research Fellow at Antenna and EMC Lab (LACE) of Istituto Superiore Mario Boella (ISMB) in Turin, Italy. From Aug 2015 to Oct 2017, he was a Postdoctoral Research Fellow at SUTD-MIT International Design Center and Division of Engineering Product Development (EPD) of Singapore University of Technology and Design (SUTD). Since November 2017, he has joined Information Technology University (ITU) of the Punjab, Lahore, Pakistan as an Assistant Professor. His current research interests are charge transport, electron device modeling, electromagnetic field theory, computational electromagnetics, bio-electromagnetics, , fractal electrodynamics and microwave imaging.
\end{IEEEbiography}
\begin{IEEEbiography}[{\includegraphics[width=1in,height=1.25in,clip,keepaspectratio]{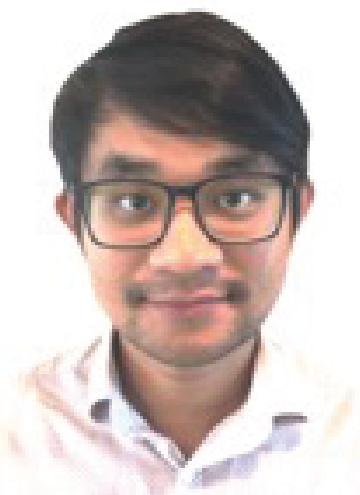}}]{Yee Sin Ang} is a research fellow at the Singapore University of Technology and Design (SUTD), Singapore. He received his bachelor’s degree in medical and radiation physics in 2010, and his PhD degree in theoretical condensed matter physics in 2014 from the University of Wollongong (UOW), Australia. He received UOW-EIS Faculty Best Postgraduate Thesis Award (2015), IPS Meeting Outstanding Poster Award (2016), and FIRST Industrial Workshop: First Prize Winner (2017). His research interests include the theory and mathematical modelling of electron emission phenomena in 2D and topological materials, electron transport physics across 2D/3D, 2D material valleytronics, nanoelectronics and superconducting devices. 
.
\end{IEEEbiography}
\begin{IEEEbiography}[{\includegraphics[width=1in,height=1.25in,clip,keepaspectratio]{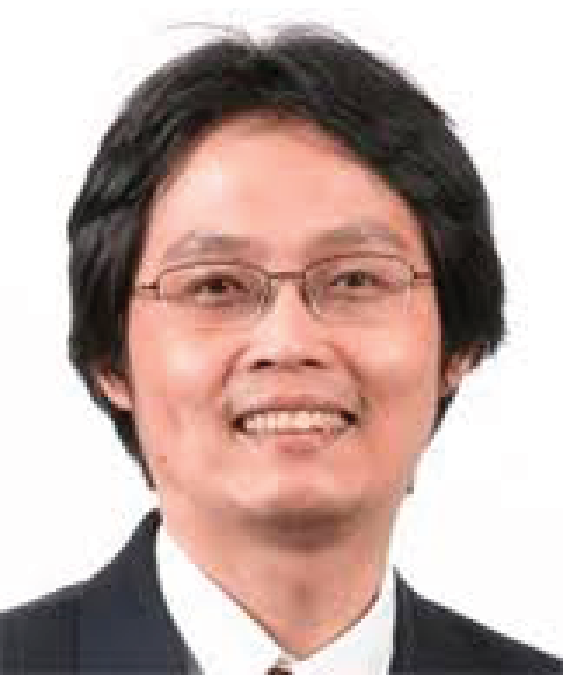}}]{Lay Kee Ang } (S'95-M'00-SM'08) received the B.S. degree from the Department of Nuclear Engineering, National Tsing Hua University, Hsinchu, Taiwan, in 1994, and the M.S. and Ph.D. degrees from the Department of Nuclear Engineering and Radiological Sciences, University of Michigan, Ann Arbor, MI, USA, in 1996 and 1999, respectively.He received a fellowship to work as a Los Alamos National Laboratory Director Postdoctoral Fellow in the Plasma Physics Applications Group, Applied Physics Division, from 1999 to 2001. He was an Assistant Professor and a tenured Associate Professor in the Division of Microelectronics, School of Electrical and Electronic Engineering, Nanyang Technological University, Singapore, from 2001 to 2011. Since 2011, he has been with the Singapore University of Technology and Design, Singapore, and is currently the Director of the Office of Graduate Studies and a Professor with the Engineering Product Development pillar. He is appointed Ng Teng Fong Chair Professor for the SUTD-ZJU Innovation, Design, and Entrepreneurship Alliance, since 2016. He has published more than 70 journal papers on these topics. His research interests include electron emission from novel materials, ultrafast laser induced photocathode, space charge limited current, multipactor discharge, plasmonics, and charge injection into solids.He was the founding Chairman of the IEEE NPSS chapter in Singapore in 2012. He received several Window of Science Awards from AFOSR-AOARD to be short-term visiting scientist to USA.
\end{IEEEbiography}

\end{document}